# Functional Group Induced Transformations in Stacking and Electron Structure in Mo$_2$CT$_X$/NiS Heterostructures

Jiamin Liu, Guo Li, Xinxu Zhang, Jiahao Wei, Hui Jia, Yulong Wu, Changlong Liu and Yonghui Li*

Department of Physics and Tianjin Key Laboratory of Low Dimensional Materials Physics and Preparing Technology, School of Science, Tianjin University, Tianjin 300350, China

Corresponding author: Yonghui Li, email: yonghui.li@tju.edu.cn

**Abstract**

The two-dimensional transition metal carbide/nitride family (MXenes) has garnered significant attention due to their highly customizable surface functional groups. Leveraging modern material science techniques, the customizability of MXenes can be enhanced further through the construction of associated heterostructures. As indicated by recent research, the Mo$_2$CT$_x$/NiS heterostructure has emerged as a promising candidate exhibiting superior physical and chemical application potential. The geometrical structure of Mo$_2$CT$_x$/NiS heterostructure is modeled and 6 possible configurations are validated by Density Functional Theory simulations. The variation in functional groups leads to structural changes in Mo$_2$CT$_x$/NiS interfaces, primarily attributed to the competition between van der Waals and covalent interactions. The presence of different functional groups results in significant band fluctuations near the Fermi level for Ni and Mo atoms, influencing the role of atoms and electron's ability to escape near the interface. This, in turn, modulates the strength of covalent interactions at the MXenes/NiS interface and alters the ease of dissociation of the MXenes/NiS complex. Notably, the Mo$_2$CO$_2$/NiS(P6$_3$/mmc) heterostructure exhibits polymorphism, signifying that two atomic arrangements can stabilize the structure. The transition process between these polymorphs is also simulated, further indicating the modulation of the electronic level of properties by a sliding operation.

## 1 Introduction

Since the first discovered MXenes (transition metal carbides, nitrides), Ti$_3$C$_2$Tx[1], many works have focused on the adjustment in structures and compositions of MXenes to achieve unique physical or chemical properties including the high conductivity; mechanical properties; functionalized surfaces ready to bond; high negative ζ potential; and effective absorption of electromagnetic waves. These reported unique properties are derived from the highly customizable/optimizable structures in MXenes. Hundreds of varieties of MXenes with different ratios of M or X elements have been synthesized via introducing stoichiometric MXenes combinations and forming of solid solutions. On the structure side, the MXenes family materials are initially built on a backbone with a general formula of M$_{n+1}$X$_n$T$_x$ (n=1, 2, 3), which is achieved via selectively etching from their parent bulk M$_{n+1}$AX$_n$ (M=early transition metal [2,3], A=elements in group IIIA/IVA, X=C/N, n=1, 2, or 3). Typically, the etchant removes A-group elements (mainly Al) and also introduces surface functional groups, T$_x$ (T$_x$=-O, -OH, -F or another group [4,5]). Besides etching, other synthesis methods, including compounding that has not been synthesized from MAX phases [6] and using chemical vapor deposition (CVD) [7,8] are also efficient ways. The diversity in synthesis techniques has led to a substantial increase in the chemical and structural complexity of MXenes.

MXenes family materials are benefit by their flexibility in structures, but such flexibility also introduces challenges. One of the challenges is the instability introduced by restacking in solvents, which can weaken their electrochemical properties, such as capacitance and conductivity. MXenes are prone to oxidation in the air, which can impact its electrochemical and mechanical properties. Therefore, designing MXenes contained heterostructures seems to resolve the drawbacks of MXenes. Inspired by heterostructures with improved electrocatalytic activity such as Ni$_3$S$_2$@MoS$_2$[9], MoS$_2$/NiS$_2$[10], MoO$_2$/Mo$_2$CT$_x$[11], and MoS$_2$-Ni$_3$S$_2$[12], MXenes/NiS heterostructures may be promising new structures with better electron structures to support electrocatalytic application. Recently, Wu et al. experimentally prepared a mixed heterostructure of $Mo_2CT_x$-MXenes and transition metal sulfides, finding that the heterostructure built with NiS and $Mo_2CT_x$-MXenes not only brought a large electrochemical surface area with abundant rich active sites exposed but also enhanced the intrinsic kinetics to promote the electrolysis process [13]. However, the atomic-level analysis of these heterostructures is still lacking, hindering further research and material design.

Among the intricate structure-activity problems in MXenes-contained heterostructures, diverse impacts caused by different $T_x$ functional groups [14–17] are still unclear. Surface functional groups affect MXenes performance in a variety of ways, such as hydrophilicity/hydrophobicity [18], stability [19], electronic conductivity [20] and magnetism [21]. Theoretical simulations by Li et al. with Density Functional Theory (DFT) show that the $Ti_3C_2T_x$/graphene heterostructure with different stacking configurations demonstrates interfacial charge transfer and adhesion energetics [22]. Zhou and colleagues report that bare $Ti_3C_2$ monolayers have low Li diffusion barriers and high theoretical Li capacities compared to the ones functionalized with -F and -OH [23]. Zhang's group reports that $Ti_3C_2$ with O- as the functional group has the best ideal strength, as evidenced by DFT calculations [24]. Therefore, it is important to theoretically investigate the NiS/$Mo_2CT_x$ heterostructure. This work aims to explore the structure of NiS/$Mo_2CT_x$ and analyze its physicochemical properties derived from the structure using DFT.

## 2 Stacking Shift Induced by Functional Groups

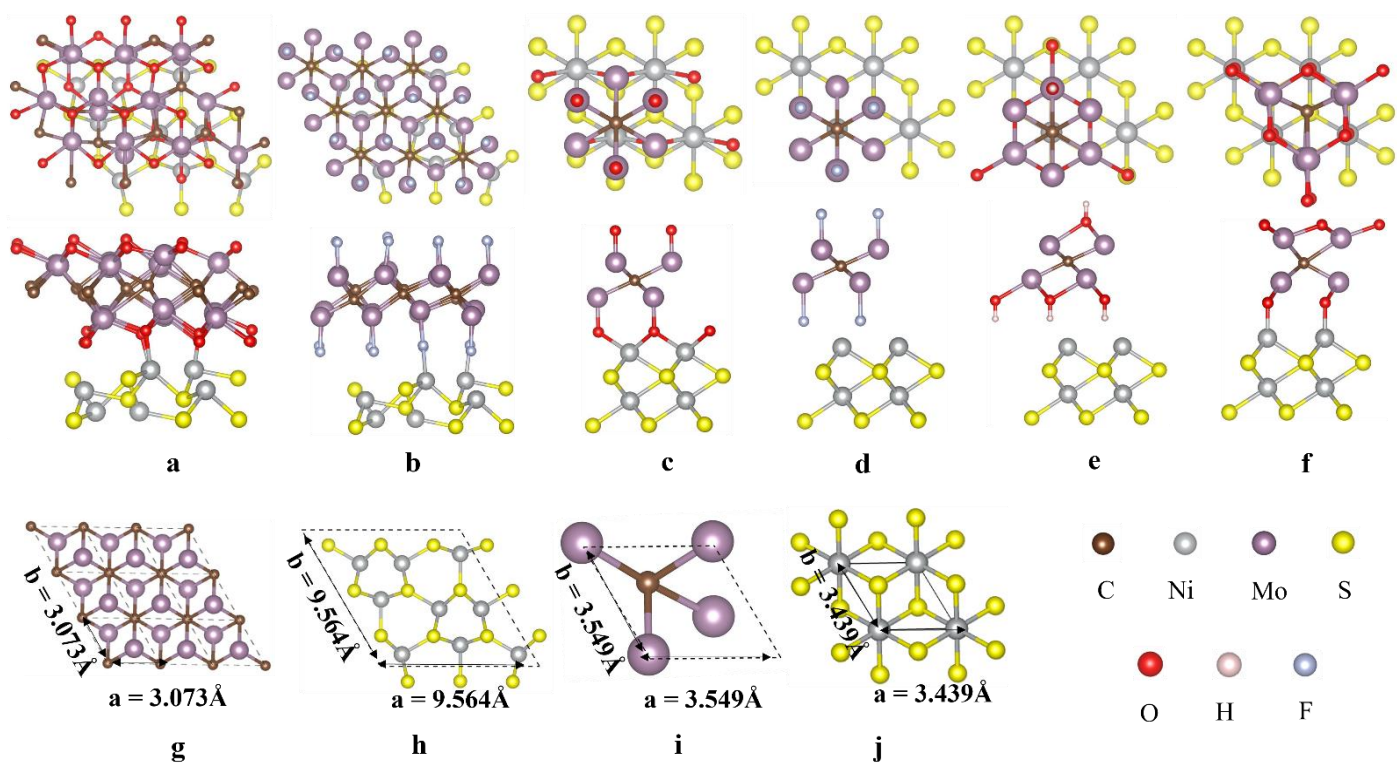

*Figure 1(a~f) the side and top views of the DFT optimized heterostructure $Mo_2CO_2$/NiS(R3m), $Mo_2CF_2$/NiS(R3m), $Mo_2CO_2$/NiS(P6$_3$/mmc)(Polymorphism I), $Mo_2CF_2$/NiS(P6$_3$/mmc), $Mo_2C(OH)_2$/NiS(P6$_3$/mmc) and $Mo_2CO_2$/NiS(P6$_3$/mmc)(Polymorphism II). (g~h) stacking of heterostructures in (a,b) without functional groups for clarity. (i~j) stacking of heterostructures in (c~f) without functional groups for clarity. In the second type of stacking, the 30° angle rotated unit cell of $Mo_2C$ matches the NiS(P6$_3$/mmc), crystal.*

The theoretical simulations begin with modeling $Mo_2CT_x$/NiS heterostructures by matching lattice structures of MXenes and NiS crystals. Considering the two crystal structures of NiS(R3m) and NiS(P6$_3$/mmc) in Figure 1h, j, the MXenes layers align with NiS crystals with 15 possibilities (**Table 1**). After DFT optimization, there are 6 confirmed configurations illustrated in Figure 1a-f: $Mo_2CO_2$/NiS(R3m), $Mo_2CF_2$/NiS(R3m), $Mo_2CO_2$/NiS(P6$_3$/mmc) (Polymorphism I), $Mo_2CF_2$/NiS(P6$_3$/mmc), $Mo_2C(OH)_2$/NiS(P6$_3$/mmc) and $Mo_2CO_2$/NiS(P6$_3$/mmc) (Polymorphism II).

In the modeling phase, many attempts are made to minimize the mismatches between the two materials. Among all attempts, successful operations can be summarized: 1) the exposed sides of NiS crystals are along their (001) direction; 2) matching NiS(R3m) with $Mo_2CT_x$ with both lattice vectors parallel respectively; 3) rotating $Mo_2CT_x$ by a 30° angle to match NiS(P6$_3$/mmc); 4) $Mo_2CT_x$ is considered to be flexible in its plane to match NiS; 5) verifying configurations with different functional groups (T=O, F, OH) via DFT. Based on the lattice constants in the literature, the mismatch of $Mo_2CT_x$/NiS(R3m) and $Mo_2CT_x$/NiS(P6$_3$/mmc) are approximately 3.73% and 3.2% respectively. Besides the lattice matching, alignment should be considered. $Mo_2CT_x$/NiS(R3m) has three possible alignments (**Table 1**): the C atom aligns with the underlying Ni atom, the S atom, or the center of any 3 neighboring Ni (or S) atoms. $Mo_2CT_x$/NiS(P6$_3$/mmc) is modeled with two possible alignments considered (**Table 1**): The C atom can align with the underlying Ni atom or the S atom. The alignments serve as initial assumptions of the structures and may be changed significantly after DFT optimization.

Spin-polarized DFT simulations are conducted using the Vienna ab initio simulation package (VASP) [25]. It is essential to ensure a vacuum region exceeding 20Å to avoid interactions between mirror images near the periodic boundaries. In the simulations, 12×12×1 or 5×5×1 Monkhorst-Pack grids [26] are used for different configurations. The interaction between electrons and nuclei is treated using the Projector Augmented Wave method (PAW) with pseudopotentials. The plane wave expansion cutoff energy is set to 520 eV, with a total energy convergence criterion of $10^{-8}$ eV and a force convergence criterion of 0.01 eV·Å$^{-1}$.

| Configuration | Initially guessed alignment | DFT confirmed |
| --- | --- | --- |
| NiS(R3m) + $Mo_2CO_2$ | C and Ni | Ture but shifted (Fig1a) |
|  | C and S | False |
|  | C and center of 3 neighboring Ni (or S) atoms | False |
| NiS(R3m) + $Mo_2CF_2$ | C and Ni | Ture but shifted (Fig1b) |
|  | C and S | False |
|  | C and center of 3 neighboring Ni (or S) atoms | False |
| NiS(R3m) + $Mo_2C(OH)_2$ | C and Ni | False |
|  | C and S | False |
|  | C and center of 3 neighboring Ni (or S) atoms | False |
| NiS(P6$_3$/mmc) + $Mo_2CO_2$ | C and Ni | Ture but shifted (Fig1c) |
|  | C and S | Ture (Fig1f) |
| NiS(P6$_3$/mmc) + $Mo_2CF_2$ | C and Ni | Ture (Fig1d) |
|  | C and S | False |
| NiS(P6$_3$/mmc) + $Mo_2C(OH)_2$ | C and Ni | Ture (Fig1e) |
|  | C and S | False |

**Table 1:** Heterostructures are constructed by combining $Mo2CT_x$ (T=O, F, OH) with NiS(R3m) or NiS(P6$_3$/mmc) crystal. Due to the different alignment methods of the C atom, 15 possible configurations were generated. After computational simulations, 6 stable configurations are identified.

Based on the summary of optimized structures, one can see the competition between the van der Waals interaction and the covalent interaction. In configurations 2, 3, 4, and 5 (**Figure 1**b-e), the covalent interaction is notably overshadowed by the van der Waals interaction from heterostructure interface since the functional groups find their interlocking positions to compress the space in the heterostructure. This arrangement results in a more compact atomic stacking, enhancing the stability of the union between the two materials. On the other hand, configurations 1 and 6 (**Figure 1**a, f) presenting a different scenario. In these configurations, the covalent interactions between functional groups and Ni atoms surpass the van der Waals interactions and leave a larger spacing at the interface. Therefore, the behavior near the heterostructure interface can be partially controlled by selecting a particular functional group: using OH or F functional groups excludes the possibility of covalent interaction and yield a heterostructure with interlocking of atoms near its interface. In the systems with O functional groups, O atoms may form bonds with Ni atoms or fill the surface space of NiS. The presence of different functional groups results in vastly different behaviors at the interface of the heterostructures.

## 3   Softness and Electron Structure Shifts

**Band structures, bonding and work functions**

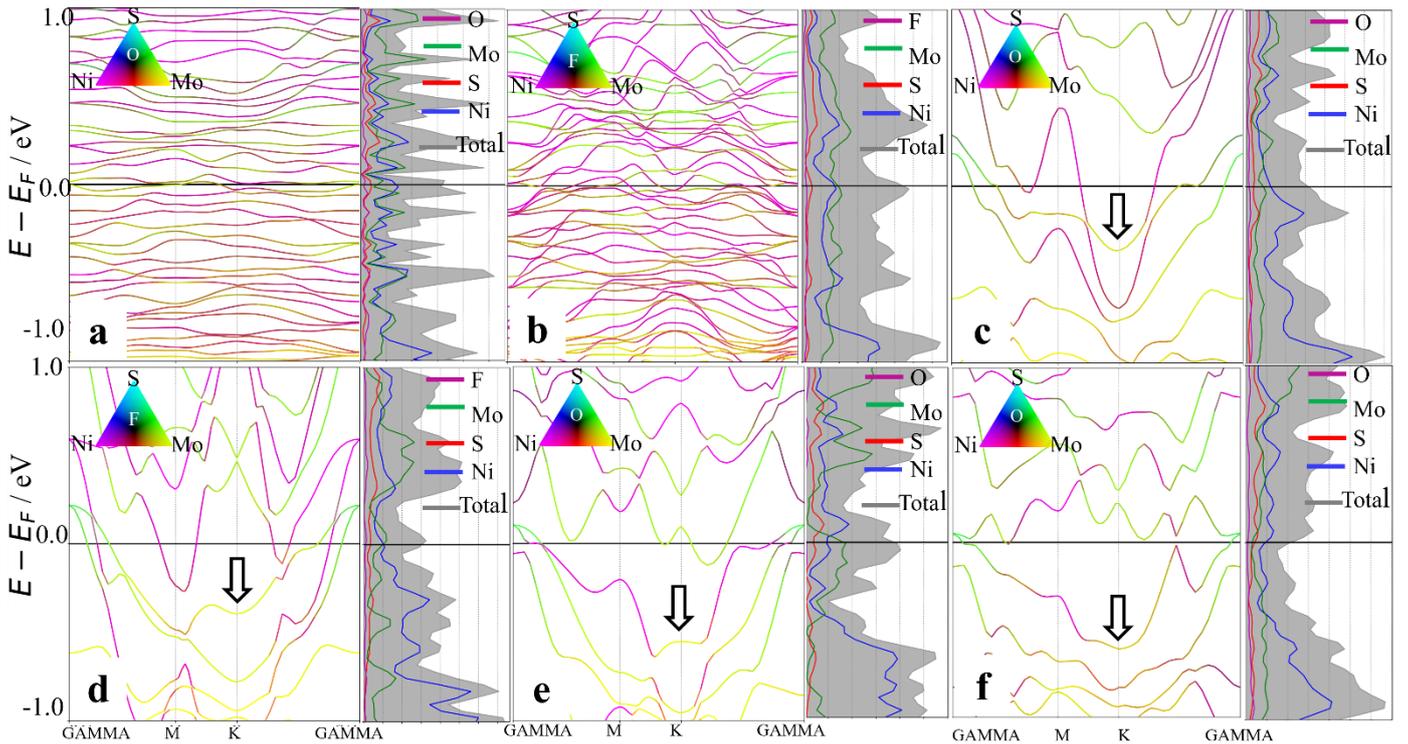

*Figure 2: (a~f) band structures and density of states plots with projected by elements for the heterostructures of the heterostructures Configuration1,2,3(Polymorphism I),4,5 and 6(Polymorphism II) respectively. The arrows in e~f are identified Mo-Ni patterns. The shape of the Mo-Ni pattern curve is a result of functional group variation.*

The structures of NiS significantly impact the overall pattern in the band structure plots with shifts contributed by MXenes. Among the 6 configurations, it is apparent that the two types of band structures are determined by NiS (**Figure 2**a-b vs. **Figure 2**c-f). Specifically, DOS plots in **Figure 2** show that the Mo and Ni atoms predominantly contribute to the electronic states near the Fermi level while the contributions from functional groups are insignificant. Thus, the conducting of NiS and MXenes maybe uncorrelated with low possibility that the current flow across the interface. Although there is negligible contribution of the functional groups to the DOS, the band structure is shifted via the geometrical impact of the functional groups. This impact can be observed by identifying characteristic patterns (indicated arrows) from the elemental projected band structures. For simplicity, the characteristic pattern along point K may be referred to as the Mo-Ni pattern. As shown in **Figure 2**c-d, the Mo-Ni pattern is located at a high level of energy while the Mo-Ni pattern can be relocated to a much lower energy level in **Figure 2**e-f. This effect of the functional groups alters the intersection between the Fermi level and the band structure, indicating a variation in the role of atoms in electrical conduction. As displayed in Configurations 3 and 4 (**Figure 2**c-d), Mo, Ni, and S all contribute to conduction. However, in Configurations 5 and 6, Mo atoms contribute less to conduction compared to Configurations 3 and 4 (**Figure 2**e-f). By contrasting the band structures and DOS plots of these heterostructures, we discern that different functional groups introduce distinct electronic characteristics near the Fermi level. The presence of these functional groups significantly modulates the conducting properties of the MXenes-based heterostructures.

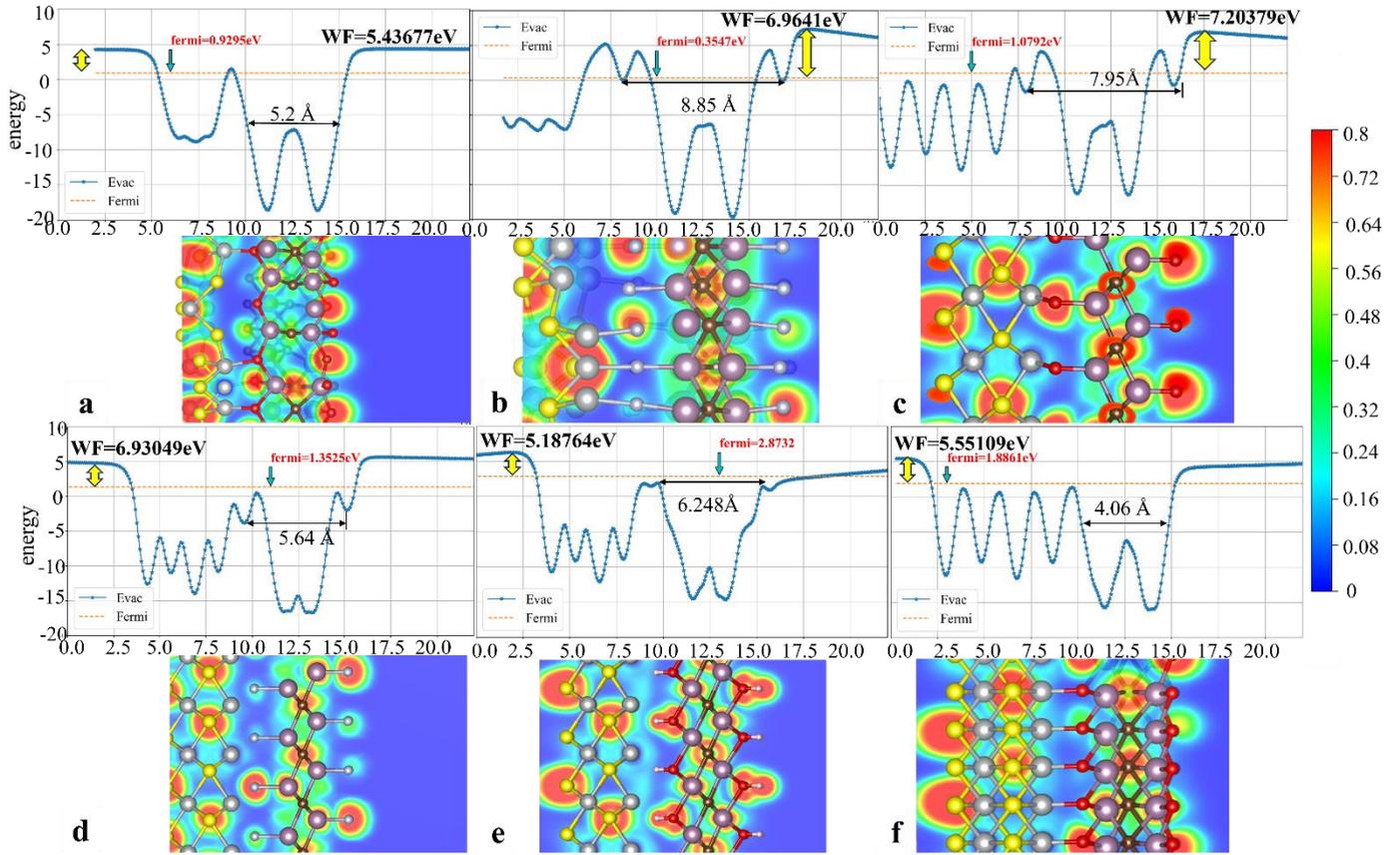

***Figure 3***: *(a,b)* work functions (upper panel) and Electron Localization Functions (lower panel) of the heterostructures Configuration1,2,3(Polymorphism I),4,5 and 6(Polymorphism II). The horizontal axis represents the position along the z-direction, and the vertical axis represents the electrostatic potential energy at different positions. According to the definition of the work function, $W_F = E_{vac} - E_{fermi}$, where Evac denotes the vacuum energy level, and $E_{fermi}$ represents the Fermi energy level. The difference between the two indicates the energy required for electrons to move from the Fermi level of the heterostructure to the free space outside the heterostructure surface, i.e., $W_F$. Regions with high probability for finding electron pairs are filled by red (ELF≥0.8).

The manner of the functional groups at the interfaces can be represented by joining or splitting peaks in the planar averaged electrostatic potential (ESP) curve (**Figure 3**a-f). In Configurations 1, 3 and 6 where the covalent interaction dominates, there are joined peaks across the interface to depict such binding. For Configurations 2, 4 and 5, peaks from functional groups and NiS surfaces are still distinctive even influenced by the van der Waals interactions.

Even in the van der Waals interaction dominated interface, the covalent interactions may still contribute to the interlocking and shift the spacing in the junction. In Configuration 2, there is a barrier (between F and NiS) close to the height of vacuum level while in Configuration 4, such barrier between F and NiS drops below the Fermi energy. In the Electron Localization Function (ELF), the ionic bonding also corroborates this observation. Compared to Configuration 4, Configuration 2 exhibits a more pronounced gap of electron density absence between the charge density regions of F and Ni atoms at the interface. Thus, the space of the interface in Configuration 4 is much smaller than that in Configuration 2. Ni-S covalent interactions in the "non-uniform" structure of NiS(R3m) partially prevent the interlocking of F atoms and NiS surface. In contrast, as a perfect hexagonal surface of NiS(P6$_3$/mmc), the interlocking happens smoothly.

Moreover, the functional groups also reshape the internal structure within MXenes layers. When one focus on the position of the C atom, functional groups F and OH behave differently from O by shifting the central C to a much lower energy in the work function curve. Such lowering of C corresponding energy on the work function curve can be associated with the thickness in the MXenes of van der Waals dominated heterostructures. Analyzing the work function of the six heterostructures, in Configurations 2, 3, 4, and 5, the thickness of MXenes increases due to the influence of van der Waals interactions. On the other hand, in Configurations 1 and 6, the covalent bonding between heterostructure layers counteracted the tendency for MXenes to thicken. Upon a detailed examination of the work functions of these six configurations and considering the O atom as a functional group, Configurations 1 and 6 exhibit notably lower work functions than Configuration 3. This suggests a greater ease of electron release from the surface of the former configurations. This behavior can be attributed to the interplay between van der Waals interactions and covalent bonding at the heterostructure interface. Notably, Configuration 5 possesses the lowest work function, underscoring its tendency to liberate electrons and its superior surface reactivity. This further highlights the pronounced sensitivity of the work function to the presence of functional groups.

## Spacing in the Heterostructures

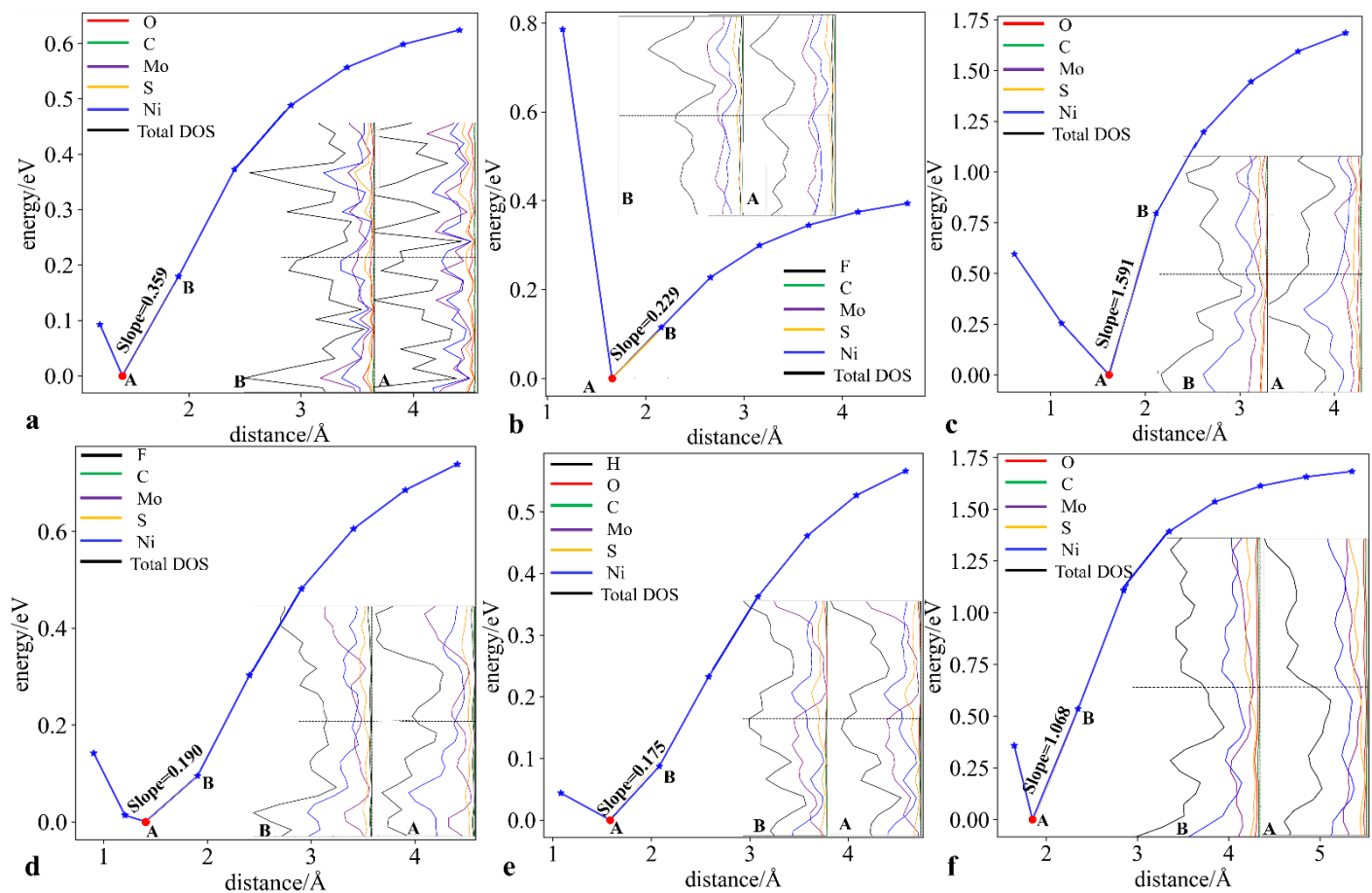

*Figure 4: The energy-space curves of the heterostructures Configuration1,2,3(Polymorphism I),4,5 and 6(Polymorphism II). As the space between $Mo_2CT_x$ and NiS increases from the optimized distance, increases in energy behave differently. Additionally, each inset shows the projected DOS (-0.5 to 0.5eV) of the heterostructures at points A (optimized structure) and B (0.5Å extension in interface spaces) in the graph respectively. Increases in energy when interface spaces are extended are also included as the slopes of the energy spacing curve.*

To further explore the competition between interactions at the interfaces, the profile of energy change of the heterostructure by varying the distance between $Mo_2CT_x$ and NiS are simulated. **Figure 4**a, c and f display the behavior of the covalent dominated interaction at the interfaces, as indicated by the energy leap when separating $Mo_2CT_x$ and NiS. In Configuration 1, the Ni-O interactions are weakened by the "non-uniform" structure of NiS(R3m) which yields a lower slope in energy increase. **Figure 4**b, d and e depict the van der Waals interaction dominated interface with a much lower slope in energy increase. Thus, in Configurations 1, 2, 4 and 5, it costs less than 1eV to dissociate the heterostructure. Configuration 3 and 6 may potentially be the target heterostructures to synthesize.

When the O atom serves as the functional group, there is a marked decrease in the DOS contribution from the Mo atom near the Fermi level as the distance between $Mo_2CO_2$ and NiS expands. Specifically, in Configuration 3, the dissociation slope is the most pronounced among all configurations. This suggests a heightened sensitivity of its covalent interactions to the interlayer distance, causing the DOS of the Ni atom to split from a primary peak into three secondary peaks. Furthermore, in Configuration 4, there is a substantial decrease in the DOS contribution from the Mo atom, resulting in the disappearance of the primary peak in the total density of states. Owing to the differential dominance of van der Waals interactions and covalent bonding, coupled with the influence of functional groups, the DOS undergoes peak splitting/vanishing near Fermi levels as the distance between $Mo_2CT_x$ and NiS varies.

## Sliding in the Heterostructures

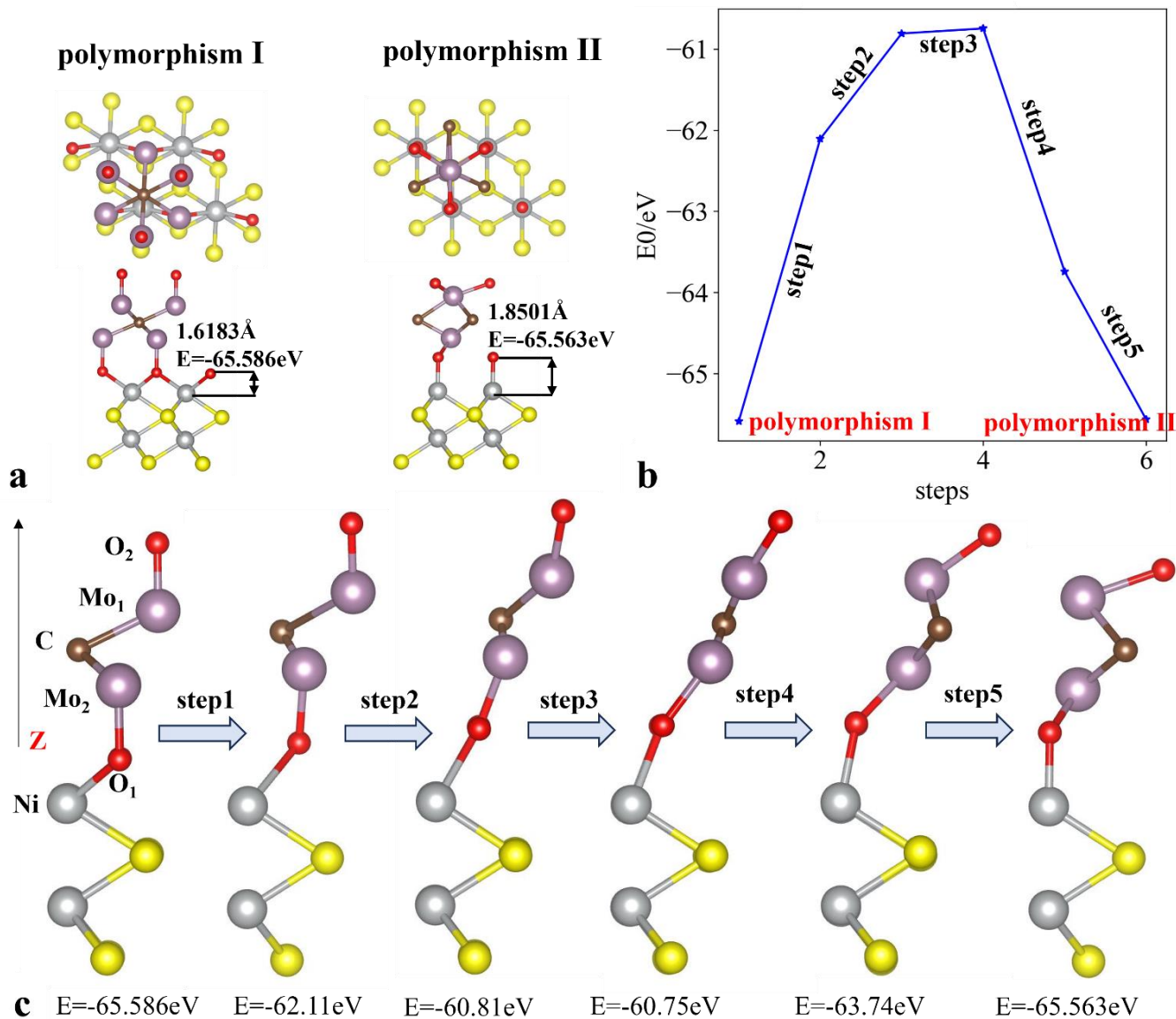

**Figure 5:** *(a) the front and top views of the two polymorphisms of $Mo_2CO_2$/NiS($P6_3$/mmc). The O atom, acting as a functional group, has two placement sites: polymorphism I aligns with the Mo atom, while polymorphism II aligns with the Ni atom. (b) the energy of each state during the transformation process between the two polymorphisms. (c) the positional changes of each atom in the two polymorphisms of $Mo_2CO_2$/NiS($P6_3$/mmc) as they transform from polymorphism I to polymorphism II. In the modeling of the translational structures, the 2 Mo atoms, 1 C atom, and 2 O atoms are numbered along the Z-direction and their positions are transcribed into relative coordinates. Throughout the calculation, the Ni atoms remain stationary.*

Considering the stability, interface binding, and low energy difference (ΔE=0.024eV), a transition can be expected from one energetically favorable structure (Configuration 3) to another equally stable arrangement (Configuration 6). Configuration 3 and 6 with details are showcased in **Figure 5**a. Given the two structures, a potential transitional pathway can be designed in following simulation steps: 1) Within each polymorphism, convert all atoms in MXenes to their relative coordinate. I.e., The O1 atom orbits with the Ni atom as its reference, Mo1 orbits O1, C orbits Mo1, Mo2 orbits C, and O2 orbits Mo2; 2) Transmute each relative coordinate into its spherical coordinate form. 3) Leverage interpolation to add 4 transitional structures for DFT energy calculations; 4) Convert the interpolated transitional structures to relative coordinates and then to laboratory coordinates for simulation. The following phase adopts the coordinates secured from the prior phase as its genesis reference for the next step.

The transition of Configuration 3 to 6 via sliding is constituted of multiple rotations with all atoms in MXenes involved. As shown in **Figure 5**c and **Table 2**, the O1 atom rotates clockwise with the Ni atom as its pivot. The Mo1 atom gyrates around O1 in a clockwise manner, C atom revolves counterclockwise around Mo1, Mo2 atom transitions clockwise around C, and O2 rotates clockwise around Mo2. The energy profile displays an energy barrier (**Figure 5**b) between Polymorphism I and II. The NiS($P6_3/mmc$) as the supporter from the heterostructure allows the sliding of MXenes from one configuration to another.

When the energy barrier can be resolved by physical/chemical techniques, the sliding does not impose any additional energy on the system but proffers the capacity to modulate its physicochemical attributes. Such sliding induced property changes can be expected. For instance, by sliding the structure from Configuration 3 to 6, a work function can be reduced from 7.20eV to 5.55eV which allows further chemical processes on such "activated" structure.

| Atom | $\triangle r$/ Å | $\triangle \theta$/Rad | $\triangle \varphi$/Rad |
|---|---|---|---|
| O1 | -0.20408333 | -0.23099366 | -0.09602352 |
| Mo1 | 0.05576439 | 0.20282898 | 0.6209008 |
| C1 | 0.01046453 | -0.00756422 | 0.23154253 |
| Mo2 | 0.01089098 | -0.03109446 | 0.43503149 |
| O2 | 0.08301939 | 0.24494919 | 0.43273539 |

*Table 2:* *The display shows the magnitude of change for each atom in spherical coordinates relative to a specified reference point at each step.*

## 4  Summary

In summary, employing DFT, 6 possible $Mo_2CT_x$/NiS heterostructures are explored with details in structures and electron properties. It is the impact from the functional groups that stems from the competition between covalent interactions and van der Waals forces at the MXenes/NiS interface. Such functional group induced competition plays an essential role in their band structures, work functions, bonding and variations in dissociation tendencies. Functional groups determine the alignment of MXenes and NiS at the interface. Thus, the roles of Mo, Ni and S are shifted as a part of whole band structure change. Differences in functional groups lead to an alternating contribution of Ni and Mo atomic energy levels near the Fermi energy. This alternation results in varying work functions within the metal heterostructures with enhanced covalent interactions at the interface. Additionally, the easier dissociation of MXenes/NiS is primarily due to the presence of O atom as the functional group. Furthermore, a transition dynamics model between two stable structures, Configuration 3 and 6 (two distinct polymorphisms of $Mo_2CO_2$/NiS($P6_3$/mmc)), is extensively explored, revealing the associated energy variations and the energy barrier. The NiS provides the foundation of the geometrical sliding of top MXenes where the van der Waals interactions secure the basis for the geometrical shifts. Such changes in geometry suggest a modulation pathway: microscopic properties (work functions, bindings and etc.) can be changed from a macroscopic manner. Beyond the pivotal advancements in the synthesis, expansion, structural and chemical understanding, as well as the properties and applications of the MXenes family, this discourse also hints at prospective research avenues in this rapidly advancing domain.

## Acknowledgement

Yonghui Li was sponsored by the National Key Research and Development Program of China (Grant No.2021YFF1200701), the National Natural Science Foundation of China (Grant No.11804248) and the Key Projects of Tianjin Natural Fund 21JCZDJC00490. Changlong Liu was supported by the National Natural Science Foundation of China (No. 11675120 and 11535008) and the National Key Research and Development Program of China (No. 2021YFF1200700).